\documentclass{an}
\usepackage{graphicx}
\usepackage{times}
\Volume{000}             
\Year{0000}              
\Month{00}               
\Pagespan{000}{000}      

\begin{document}
\lhead[\thepage]{A.N. H. Eerik, P. Tenjes: ~Metallicity distribution of GCSs}
\rhead[Astron. Nachr./AN~{\bf 3XX} (2002) X]{\thepage}
\headnote{Astron. Nachr./AN {\bf 3XX} (2002) X, XXX--XXX}

\title{Metallicity distributions of globular cluster systems in galaxies}

\author{H. Eerik$^1$ \and P. Tenjes$^{1,2}$}
\institute{Institute of Theoretical Physics, Tartu University,
   T\"ahe 4, Tartu, 51010 Estonia \and  Tartu Observatory, T\~oravere 
   61602, Estonia}
\date{Received {\it date will be inserted by the editor}; 
accepted {\it date will be inserted by the editor}} 

\abstract{
	We collected a sample of 100 galaxies for which different observers have 
determined colour indices of globular cluster candidates. The sample includes 
representatives of galaxies of various morphological types and different 
luminosities. Colour indices (in most cases $(V-I)$, but also $(B-I)$ and 
$(C-T_1)$) were transformed into metallicities $\rm [Fe/H]$ according 
to a relation by Kissler-Patig (1998). These data were analysed with the KMM 
software in order to estimate similarity of the distribution with uni- or 
bimodal Gaussian distribution. We found that 45 of 100 systems have bimodal 
metallicity distributions. Mean metallicity of the metal-poor component 
for these galaxies is ${\rm \langle [Fe/H]\rangle} = -1.40 \pm 0.02,$ of the 
metal-rich component ${\rm \langle [Fe/H]\rangle} = -0.69 \pm 0.03$. Dispersions 
of the distributions are 0.15 and 0.18, respectively. Distribution of unimodal 
metallicities is rather wide. These data will be analysed in a subsequent paper 
in order to find correlations with parameters of galaxies and galactic environment.
}
\correspondence{ptenjes@ut.ee}

\maketitle

\section{Introduction}

In galaxy formation theories different physical processes may determine the 
formation and evolution of galaxies with different luminosities and different 
morphological types. Star formation is regulated by hierarchical clustering of 
dark matter halos, accretion of gas into dark matter potential well, reionisation 
of ISM by first stars, outflows of gas, galactic mergers, etc. On the other hand, 
despite varieties between galaxies it is important to find similarities 
between them. Globular cluster systems (GCSs) are known to exist in galaxies 
with rather different morphology (see e.g. Ashman \& Zepf 1998). Therefore, it 
is interesting to compare the properties of GCSs in galaxies with different 
morphological types and different luminosities. 

It has been firmly established that GCSs of many galaxies may have bimodal 
metallicity distribution (e.g. Ashman \& Zepf 1998; Gebhart \& Kissler-Patig 1999),
and different scenarios have been proposed to explain bimodality. It is rather 
reasonable that mergers between galaxies may stimulate the formation of globular
clusters (GCs) as well as dwarf galaxies (Schweizer 1987; Ashman \& Zepf 1992; 
Weilbacher et al. 2000), but before accepting mergers as a dominating mechanism 
in creating bimodality, additional studies and arguments are needed. A multi-phase 
collapse process (Forbes et al. 1997) as a dominating mechanism needs additional
theoretical calculations based on the general theory of the reionisation epoch in 
galactic formation. Accretion of metal-poor GCs from dwarf galaxies (C\^ot\'e et 
al. 1998) is also a rather plausible process, but before accepting it as a dominating 
process, additional studies are needed. Although the reasons of uni- or
bimodality have not been fully understood, serious attempts have been made to 
explain bimodality in the case of some particular galaxies or particular group of 
galaxies (e.g. van den Bergh 1998; Harris et al. 2000; Forbes et al. 2001; 
Larsen et al. 2001a; Ashman \& Zepf 2001).

In this paper, we compile a sample of colour distributions of GCS in galaxies. 
Colours were corrected from absorption in our Galaxy according to Schlegel et al.
(1998) if not mentioned otherwise. Colours were analysed with the help of the KMM 
algorithm (Ashman, Bird \& Zepf 1994). As initial data for the KMM we used individual 
cluster data and not the histograms of metallicity distributions. The KMM algorithm 
allows to determine the maxima in uni- or bimodal metallicity distributions. 
Correlations of derived metallicities with various properties of host galaxies and 
spatial distribution parameters of GCSs will be studied in a subsequent paper.

\section{Sample and metallicity calibration}

Colour distributions are now available for GCSs in nearly 110 galaxies. 
However, in some galaxies the observed number of clusters is too small for 
statistical analysis. After rejecting these systems, our sample includes GCSs in 
100 galaxies. This sample contains representatives of host galaxies of different 
morphological types and different luminosities. Tables~\ref{gendata1} and 
\ref{gendata2} present the names, morphological types, absolute luminosities 
in V colour, estimated total numbers of globular clusters and references to 
the sources of globular cluster data. 

\begin{table}[ht]
\caption{General data about galaxies with GCSs}
\label{gendata1}
\begin{tabular}{llccl}\hline
Galactic & Morph. & $M_{V}$ & $N$           & Source \\ 
NGC name & type   &         &               &        \\ 
\hline
224      & Sb     & -21.8   & $450\pm 100$  & 2 \\ 
474      & S0     & -21.4   & $\sim 500$    & 1 \\ 
524      & S0     & -22.1   & $3300\pm 1000$& 2 \\ 
584      & E4     & -21.4   & $406\pm 52$   & 3 \\ 
596      & Ep     & -20.9   &               &  \\ 
598      & Scd    & -19.2   & $75\pm 14$    & 15 \\ 
821      & E4/6   & -20.8   & $395\pm 94$   & 3 \\ 
1023     & S0     & -21.0   &               &  \\ 
1052     & E4     & -20.9   & $430\pm 80$   & 2 \\ 
1199     & E3     & -22.0   &               &  \\ 
1201     & S0     & -20.8   & $\sim 460$    & 1 \\ 
1332     & S0     & -21.2   & $400\pm 50$   & 1 \\ 
1374     & E1     & -19.8   & $410\pm 82$   & 3 \\ 
1379     & E0     & -19.9   & $314\pm 63$   & 3 \\ 
1380     & S0     & -21.7   & $560\pm 30$   & 3 \\ 
1387     & S0     & -20.2   & $389\pm 110$  & 3 \\ 
1389     & S0     & -19.5   & $\sim 60$     & 1 \\ 
1399     & cD/E1  & -21.8   & $5940\pm 570$ & 4 \\ 
1400     & S0     & -20.6   & $922\pm 280$  & 3 \\ 
1404     & E1     & -21.4   & $880\pm 140$  & 3 \\ 
1426     & E4     & -20.4   &               &  \\ 
1427     & E5/3   & -20.4   & $510\pm 87$   & 3, 14 \\ 
1439     & E1     & -20.4   & $223\pm 24$   & 3 \\ 
1553     & S0     & -21.2   & $600\pm 130$  & 3 \\ 
1700     & E4     & -22.2   & $1320\pm 270$ & 5 \\ 
2300     & S0     & -20.8   &               &  \\ 
2434     & E0     & -20.6   &               &  \\ 
2768     & S0     & -21.9   & $\sim 1000$   & 1 \\ 
2778     & E2     & -19.4   &               &  \\ 
2902     & S0     & -20.2   & $\sim 70$     & 1 \\ 
3031     & Sab    & -21.1   & $210\pm 30$   & 2 \\ 
3056     & S0     & -18.9   & $\sim 45$     & 1 \\ 
3115     & S0     & -20.8   & $630\pm 150$  & 3 \\ 
3311     & cD/E0  & -22.3   & $12400\pm 5000$& 3\\ 
3377     & E5/6   & -19.9   & $240\pm 60$   & 2 \\ 
3379     & E1     & -20.7   & $300\pm 160$  & 2 \\ 
3384     & S0     & -20.1   & $130\pm 70$   & 2 \\ 
3414     & S0p    & -21.0   & $\sim 750$    & 1 \\ 
3489     & S0     & -19.6   & $\sim 100$    & 1 \\ 
3585     & E7     & -21.7   &               &  \\ 
3599     & S0     & -18.4   & $\sim 45$     & 1 \\ 
3607     & S0     & -20.7   & $800\pm 560$  & 3 \\ 
3608     & E2     & -20.7   & $307\pm 41$   & 2 \\ 
3610     & E5     & -21.5   & $856\pm 460$  & 2 \\
3640     & E3     & -21.8   &               &  \\ 
3923     & E3/4   & -22.1   & $4300\pm 100$ & 3 \\ 
4125     & E6p    & -22.1   &               &  \\ 
4192     & Sab    & -20.7   &               &  \\ 
4203     & S0     & -20.2   & $\sim 240$    & 1 \\ 
4278     & E1/2   & -19.8   & $1000\pm 300$ & 2 \\ 
4291     & E2     & -20.7   & $714\pm 126$  & 2 \\ 
4343     & Sb     & -18.8   &               &  \\ 
4365     & E3     & -21.8   & $2500\pm 200$ & 3 \\ 
4374     & E1     & -21.7   & $3040\pm 400$ & 3 \\ 
4379     & S0p    & -19.6   & $\sim 80$     & 1 \\ 
4406     & E3/S0  & -21.8   & $3350\pm 400$ & 3 \\ 
4450     & Sab    & -22.3   &               &  \\ 
4458     & E0/1   & -19.5   & $38\pm 11$    & 6 \\ 
4459     & S0     & -20.9   & $\sim 450$    & 1 \\ 
4472     & E2     & -22.6   & $6300\pm 1900$& 2 \\ 
\hline
\end{tabular}
\end{table}

\begin{table}[ht]
\caption{Continued}
\label{gendata2}
\begin{tabular}{llccl}\hline
Galactic & Morph. & $M_{V}$ & $N$           & Source \\ 
name     & type   &         &               &        \\ 
\hline
4473     & E5     & -20.9   & $217\pm 7$    & 6 \\ 
4478     & E2     & -19.9   &               &  \\ 
4486     & cD/E0  & -22.4   & $13000\pm 500$& 2 \\ 
4486B    & cE0    & -17.8   & $113\pm 4$    & 6 \\ 
4494     & E1/2   & -21.0   & $1000\pm 350$ & 2 \\ 
4526     & S0     & -21.4   & $2700\pm 400$ & 2 \\ 
4536     & Sbc    & -21.7   &               &  \\ 
4550     & S0     & -19.5   & $92\pm 5$     & 6 \\ 
4552     & E0/5   & -21.2   & $\sim 2400$   & 3 \\ 
4565     & Sb     & -21.4   & $180\pm 20$   & 7 \\ 
4569     & Sab    & -21.8   & $900\pm 300$  & 2 \\ 
4589     & E2     & -21.2   & $789\pm 123$  & 6 \\ 
4594     & Sa     & -22.2   & $1150\pm 570$ & 12 \\ 
4621     & E5     & -21.3   & $1900\pm 400$ & 2 \\ 
4649     & E2     & -22.3   & $5100\pm 160$ & 3 \\ 
4660     & E6     & -19.2   & $176\pm 6$    & 6 \\ 
4874     & cD/E0  & -23.1   & $9200\pm 1500$&  8\\ 
4881     & E0     & -21.6   & $400\pm 80$   & 2 \\ 
4889     & E5/4   & -23.5   & $17300\pm 3000$& 2 \\ 
5018     & E3     & -22.6   & $1200\pm 500$ & 4 \\ 
5061     & E0     & -21.6   &               &  \\ 
5128     & E0p    & -22.0   & $1700\pm 0.6$ & 3 \\ 
5322     & E3/4   & -22.1   & $2352\pm 1387$& 6 \\ 
5813     & E1/2   & -21.6   & $2400\pm 600$ & 2 \\ 
5845     & E      & -19.3   & $134\pm 21$   & 6 \\
5846     & E0     & -22.1   & $3120\pm 1850$& 3 \\ 
5907     & Sc     & -21.2   & $180\pm 20$   & 7 \\ 
5982     & E3     & -21.8   & $3670\pm 1000$& 6 \\ 
6702     & E3     & -21.9   & $1776\pm 926$ & 9 \\ 
6703     & S0     & -21.8   & $\sim 800$    & 1 \\ 
6861     & S0     & -21.8   & $\sim 2000$   &  \\ 
6868     & E      & -22.1   &               & 13 \\ 
7192     & S0     & -21.5   &               &  \\ 
7332     & E2     & -20.1   &               &  \\ 
7457     & E/S0   & -19.5   & $178\pm 75$   & 10 \\ 
7619     & E2     & -22.1   &               &  \\ 
7626     & E1p    & -21.9   & $1204\pm 879$ & 6 \\ 
IC1459   & E3     & -21.4   & $526\pm 59$   & 6 \\ 
IC4051   & E2     & -21.9   & $6180\pm 1100$& 11\\ 
MW       & Sbc    & -21.3   & $180\pm 20$   & 2 \\ 
\hline
\end{tabular}
Sources for total number of GCs: 1 -- Kundu \& Whitmore (2001b); 2 -- Ashman \&  
Zepf (1998); 3 -- Kissler-Patig (1997); 4 -- van den Bergh (1998); 5 -- Brown et al. 
(2000); 6 -- Kundu \& Whitmore (2001a); 7 -- Kissler-Patig et al. (1999); 8 -- 
Harris et al. (2000); 9 -- Georgakakis et al. (2001); 10 -- Chapeton et al. 
(1999); 11 -- Woodworth \& Harris (2000); 12 -- Larsen et al. (2001b); 13 -- Da 
Rocha et al. (2002), 14 -- Forte et al. (2001), 15 -- Chandar et al. 2001.
\end{table}

Although the most frequent morphological type is elliptical, the 
sample includes also spirals from Sa to Sc. Luminosities differ as much as 2 orders 
of magnitudes.

\begin{figure}[tbp]
\resizebox{\hsize}{!}{\includegraphics*{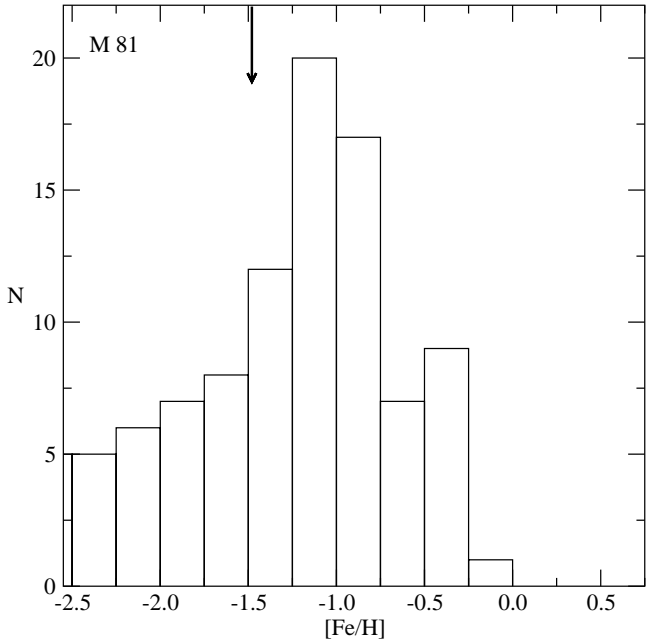}}
\caption{Metallicity distribution of M 81 GCS as an example of unimodal distribution.
Metallicity distribution is asymmetric and broad, thus the peak from KMM test (position 
of the arrow) does not coincide with the histogram maximum. Perhaps this is an example
when $(V-I)$ is not sufficiently sensitive to indicate intrinsic bimodality in 
metallicities.}
\label{f1label}
\end{figure}

\begin{figure}[tbp]
\resizebox{\hsize}{!}{\includegraphics*{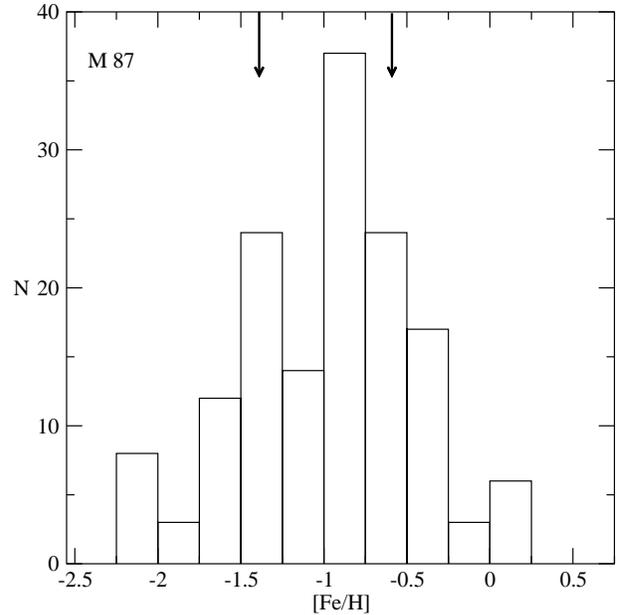}}
\caption{Metallicity distribution of M 87 GCS as an example of bimodality. Two 
arrows at the top of the figure indicate the peaks of two Gaussians from KMM test. There is 
slight asymmetry in the metal-rich peak.}
\label{f2label}
\end{figure}

\begin{figure}[tbp]
\resizebox{\hsize}{!}{\includegraphics*{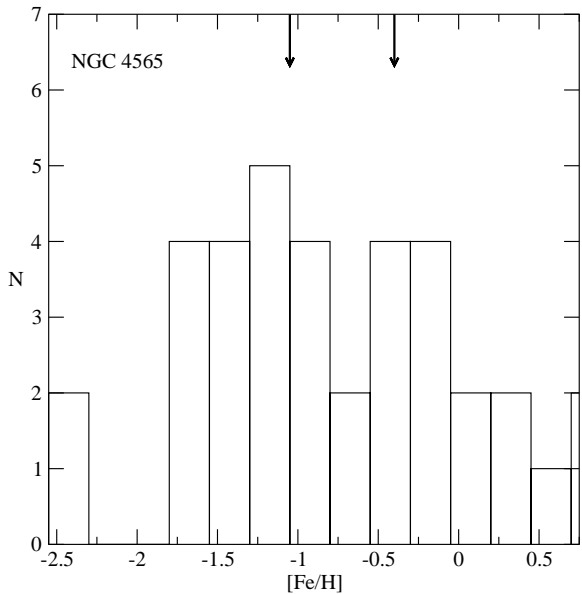}}
\caption{Metallicity distribution of NGC 4565 GCS. Another example of bimodal 
metallicity distribution. There is slight asymmetry in the metal-poor peak.}
\label{f3label}
\end{figure}

\begin{figure}[tbp]
\resizebox{\hsize}{!}{\includegraphics*{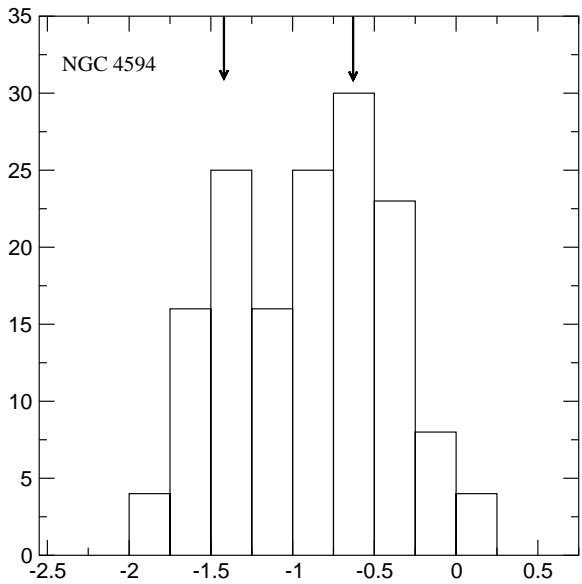}}
\caption{Metallicity distribution of NGC 4594 GCS. Good example of bimodal and 
symmetric metallicity distribution.}
\label{f4label}
\end{figure}

References to color measurements of globular clusters used in present study are 
given in Tables 3 and 4. When for a particular galaxy colors are determined by 
different studies we preferred HST observations to ground based observations 
and later observations to earlier observations. 

The next step is to move from broadband colours to metallicities. This must be 
done carefully as many calibrations are crude and small uncertainties in GC 
colours may cause large errors in derived metallicities. Calibration uncertainties 
are in most cases due to an inadequate database for calibrating any colour indices for 
metallicities above solar abundance. Nearly all the most often used empirical 
relations have been derived from colours of Galactic GCs. Due to the fact that the 
bulk of Galactic GCs are rather blue, these relations hold well only in the range of 
metallicities $\rm -2.25 \le {\rm [Fe/H]} \le -0.25.$ Extrapolating these relations 
to higher values of metallicity, they may not describe reality sufficiently well.

The relationship between the colours $(B-V)$, $(B-R)$, $(V-I)$ etc., and 
metallicities of GCs is usually approximated as linear (see e.g. Couture et al. 
1990a,b; Reed et al. 1994; Harris 1996; Kundu \& Whitmore 1998; Barmby et al. 2000). 
For the reason explained above, most of these relations overestimate the metallicity 
of red clusters. The metallicities of blue clusters are less affected by the choice 
of the transition relation (different relations give a discrepancy $\sim 0.1$ dex in 
metallicities of blue clusters, for red clusters it is $\sim 0.5$ dex). Theoretical 
models (e.g. Worthey 1994) attribute nonlinear behaviour to the relation between 
colours and metallicities, i.e. to higher metallicity values (red colours) 
corresponds a relation with a shallower slope than for blue colours. The relation 
varies also with the average age of GCS (Schulz et al. 2002).
 
With the help of more metal-rich GCs in NGC 1399 a good conversion from $(V-I)$ colour 
to super-solar metallicities has been derived by Kissler-Patig et al. (1998)
\begin{equation}
{\rm [Fe/H]} = 3.27 (V-I) - 4.50. 
\end{equation}
Although this relation is also linear, it is a good compromise to characterize 
nonlinearity of the theoretical model prediction.
Investigation of calibrations that are available for other colours shows that 
also the relation $(C-T_1)$ vs $\rm [Fe/H]$ from Geisler \&
Forte (1991)
\begin{equation}
{\rm [Fe/H]} = 2.35 (C-T_1) - 4.39 
\end{equation}
offers reliable results.

Analyses of GC metallicities have often been made using the same photometric 
filters to avoid systematic effects in the conversion to a homogeneous system. 
Today the situation is even more favourable for large sample statistics. The bulk of 
the recent GC colours available in literature come from the HST studies and often 
use the $V$ and $I$ filters (e.g. Kundu \& Whitmore 2001a,b; Larsen et al. 2001a). 
Many giant galaxies are thoroughly re-investigated in $(V-I)$. Therefore, it is 
possible to compile a homogeneous sample of systems with $(V-I)$ and $(C-T_1)$ 
colour distributions. From observations in other colour systems only GCSs in 
NGC~1316 and NGC~1404 are included in the analysis.  

Forbes \& Forte (2001) have confirmed the relations $(V-I)$ vs other colours based 
on Galactic GCs. They show that those relations are applicable for the GCs of early 
type galaxies using an extrapolation to redder colours. For NGC~1316 and NGC~1404 
GCs we use first the following relation from Forbes \& Forte (2001)
\begin{equation}
(V-I) = 0.51 (B-I) + 0.11.
\end{equation}
This relation allows to estimate $(V-I)$ with an accuracy of $0.03^m$. Using 
expression (1), we convert the derived $(V-I)$ colours to $\rm [Fe/H]$.  

\section{Results: globular cluster metallicity distributions}

In our study we use for Galactic reddening corrected colour data. Using relations 
(1) -- (3) the derived GC metallicity distributions are tested for bimodality. 

In principle, it is possible to determine the bimodality also from a classical 
histogram. However, in this study we use the KMM algorithm and unbinned GC list to 
explore bimodality. The KMM test (Ashman, Bird \& Zepf 1994) provides a statistical 
test for comparing the likelihood of the underlying distribution being a single 
or a double Gaussian. The KMM test also determines the peak values (i.e. mean 
metallicities) of each subpopulation. The peaks can be determined with an accuracy 
of 0.05 dex. Homoscedastic fitting mode was used, forcing identical dispersions. 
Only for galaxies NGC~1399 and NGC~5846 it has been determined that red GCs have 
a colour distribution almost twice as broad as the blue ones.

The input of the KMM algorithm includes individual data-points, the number of 
Gaussian components to be fitted and the starting point for the components' 
peak values and dispersions. 

Because there does not exist general databasis for globular cluster colors in 
external galaxies we give in Tables 3 and 4 references only to original papers.
For galaxies NGC 598, 1380, 1399, 3031, 4472, 4486, 4565, 4594, 5128, 5846 and 7457
data are taken from tables or figures of corresponding original papers. For galaxies
NGC 1374, 1379 and 1387 data are taken from electronic database 
(www.eso.org/~mkissler/Archive). For galaxies NGC 596, 821, 1404, 1426, 1700,2300, 
2434, 2788, 3115, 3311, 3585, 3608, 3640, 3923, 4125, 4379, 4406, 4458, 4478, 4621 
and 7192 the data are communicated by authors of corresponding papers as personal 
communication.

Since KMM is sensitive to highly deviating metallicity values in the data set, 
only the objects that are within the range $-2.55 < {\rm [Fe/H]} < 0.75$ are 
considered. The KMM test also allows to detect bimodality when the metallicity 
distributions do not visually express several peaks. In Figs.~\ref{f1label} 
-- \ref{f4label} some characteristic examples of GCSs are given. Arrow(s) at 
the top of the figures indicate the position of the central values of Gaussian(s). 
In Table~\ref{bimodal}, mean metallicities for two Gaussian components are 
given for galaxies where the KMM test indicated bimodality. In the Table the 
column labeled ``Refer." refers to the authors of GC colour measurements in a 
particular galaxy. If the KMM test was done by these authors and the methods 
correspond to ours, we did not reanalyse the corresponding data and simply accepted 
metallicity peak values. These cases are designated in the column ``Source" by 
references to corresponding authors. An asterisk in the column ``Source" means that 
the data for a particular galaxy were reanalysed by us. In Table~\ref{unimodal} 
mean metallicities for a single Gaussian component are given for galaxies where 
the KMM test discovered unimodality. 

\begin{figure}[tbp]
\resizebox{\hsize}{!}{\includegraphics*{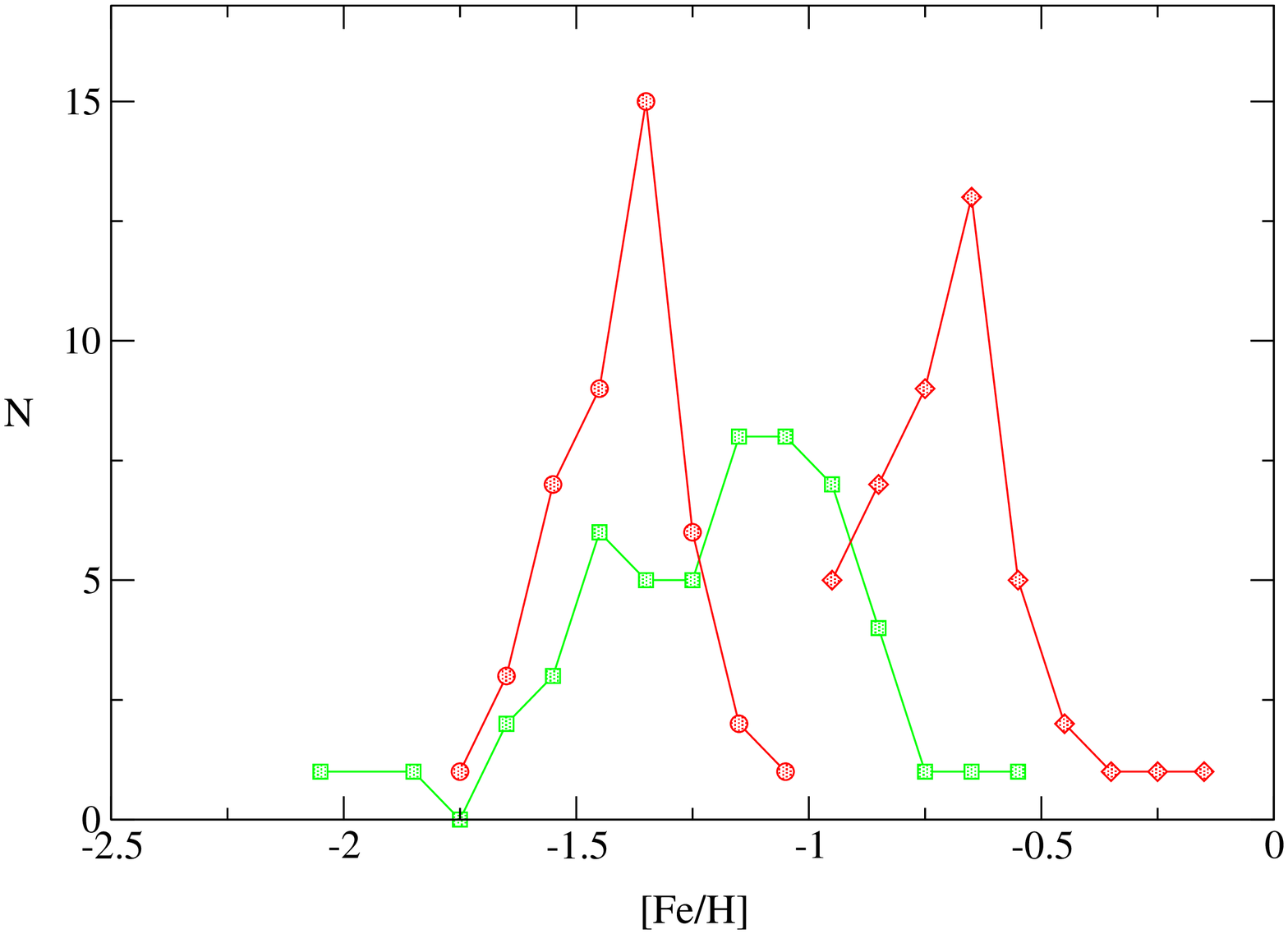}}
\caption{Distribution of metallicities of GCS. Squares -- unimodal systems, 
circles -- the metal-poor component of bimodal systems, diamonds -- the metal-rich 
component of bimodal systems.}
\label{f5label}
\end{figure}

\begin{figure}[tbp]
\resizebox{\hsize}{!}{\includegraphics*{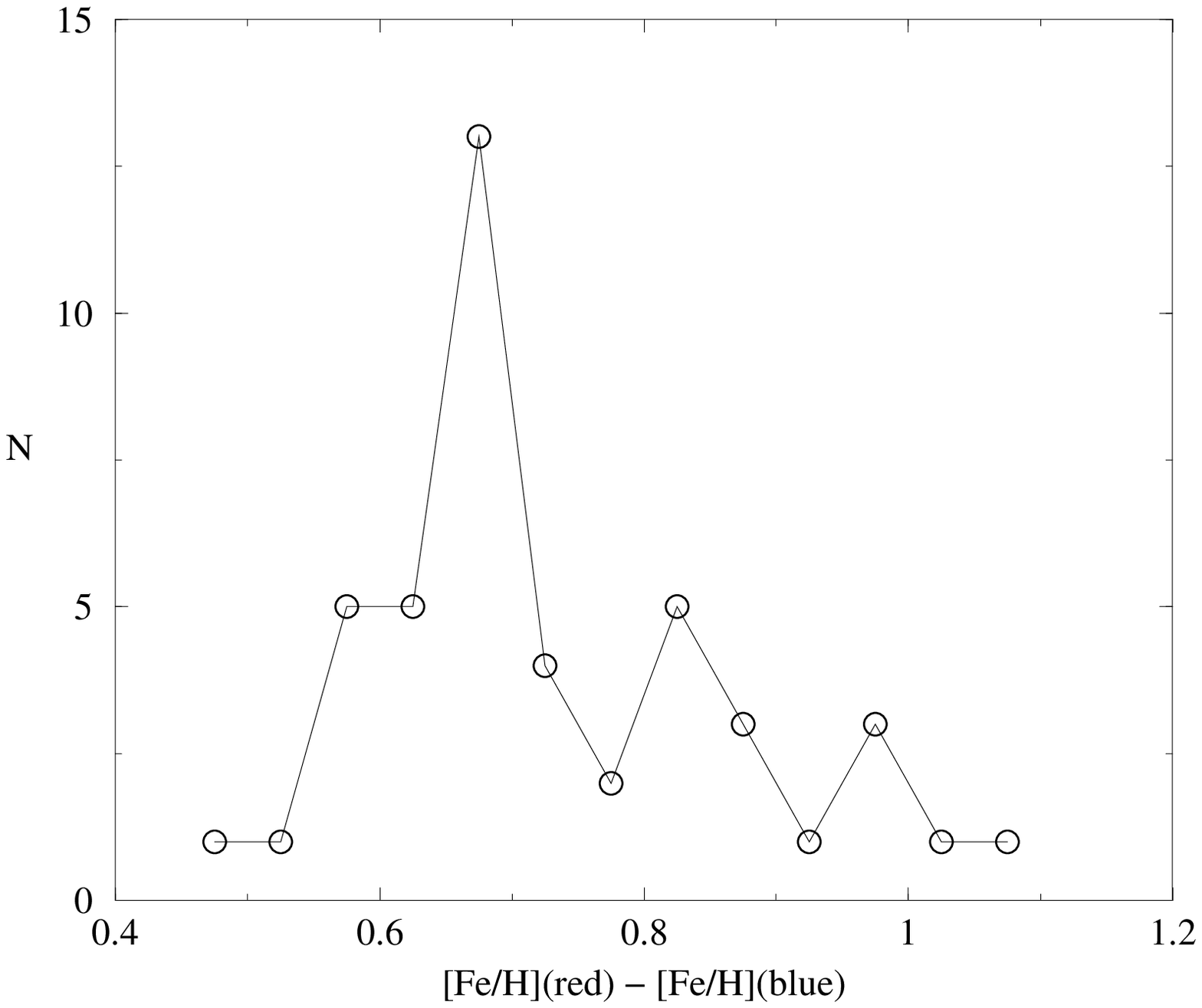}}
\caption{Distribution of differences between red (metal-rich) and blue (metal-poor) 
peaks for GCS with bimodal metallicity distribution.}
\label{f6label}
\end{figure}

\begin{figure}[tbp]
\resizebox{\hsize}{!}{\includegraphics*{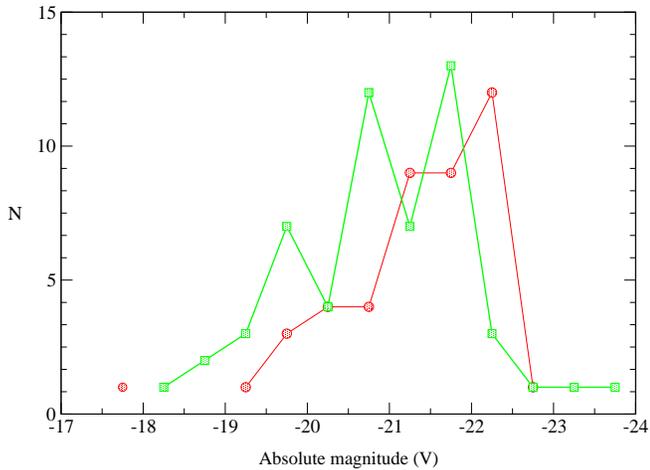}}
\caption{Distribution of absolute magnitudes of galaxies. Circles -- galaxies 
with bimodal metallicity distribution of GCs, squares -- galaxies with unimodal 
metallicity distribution of GCs.}
\label{f7label}
\end{figure}

Figure~\ref{f5label} presents the distribution of metallicities from 
Tables~\ref{bimodal} and \ref{unimodal}. Mean metallicity of the metal-poor component 
is ${\rm \langle [Fe/H]\rangle} = -1.40 \pm 0.02,$ the standard deviation of the 
distribution is 0.15. Mean metallicity of the metal-rich component is ${\rm \langle 
[Fe/H]\rangle} = -0.69 \pm 0.03,$ the standard deviation is 0.18. Both distributions 
have similar dispersions and slight asymmetry. Distribution of unimodal metallicities 
is rather wide. Distribution of metallicity differences between metal-rich and 
metal-poor peaks in galaxies is given in Fig.~\ref{f6label}. Mean difference is 
$\rm <\Delta [Fe/H]> = 0.73 \pm 0.02,$ the standard deviation is 0.14. Distribution 
is asymmetric (skewness $=0.64$), thus the median $\rm \Delta [Fe/H]=0.66$ is 
perhaps better characteristic of the distribution than the mean. Although the 
decrease of $\rm \Delta [Fe/H]$ at small values may be caused by intrinsic 
characteristics of the KMM test (smaller differences fuse into unimodal distribution) 
a steep decrease of $\rm \Delta [Fe/H]$ at higher values seems to be a real property 
of GCSs. In Fig.~\ref{f7label}, we give the luminosity distributions of galaxies with 
bimodal and unimodal metallicity distributions. In general, luminosity distributions 
have similar dispersions and skewnesses, but the galaxies with unimodal metallicity 
distribution are systematically fainter by about $-0.40$ mag.

\begin{figure}[tbp]
\resizebox{\hsize}{!}{\includegraphics*{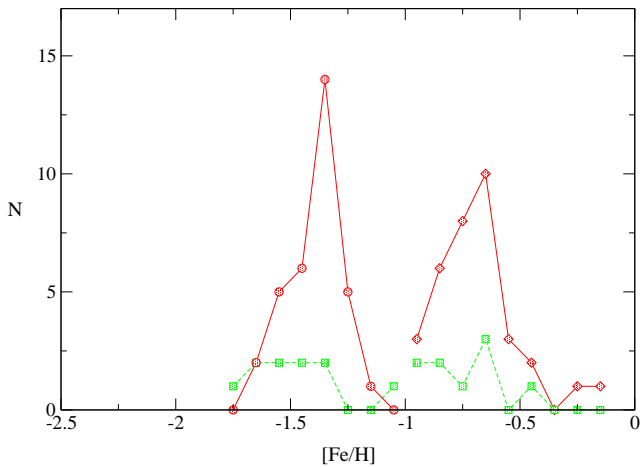}}
\caption{Distribution of metallicities of GCS, determined on the basis of
(V--I) and (B--I) colors. Squares and the dashed line -- metallicities determined 
from (B--I) colors, diamonds and the solid line -- metallicities determined from 
(V--I) colors.}
\label{f8label}
\end{figure}

\begin{figure}[tbp]
\resizebox{\hsize}{!}{\includegraphics*{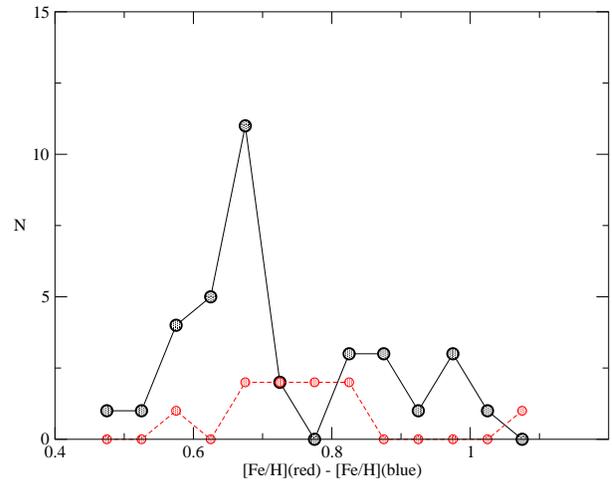}}
\caption{Distribution of differences between red (metal-rich) and blue (metal-poor) 
peaks for GCS with bimodal metallicity distribution determined on the basis of
(V--I) and (B--I) colors. The dashed line -- metallicities determined from (B--I) 
colors, the solid line -- metallicities determined from (V--I) colors.}
\label{f9label}
\end{figure}

To understand possible systematic biases in further analysis due to the
used database, we have repeated Figs.~\ref{f5label}-\ref{f6label} separately for GCSs,
metallicities of which were determined on the basis of (V--I) and (B--I) colors. $\rm
(C-T_1)$ color is used only in the case of a few galaxies and it is not possible to study
corresponding distributions. Results are presented in Figs.~\ref{f8label}-\ref{f9label}.
Although the number of (B--I) color measurements is rather small for accurate statistics 
(10 bimodal systems and only 3 unimodal systems), inspection of the figures indicates 
that metallicities determined from (B--I) colors are slightly shifted in the direction 
of smaller metallicities. This may raise a question about the precision of 
color-metallicity calibration. However, the (B--I) sample is small and thus statistical
fluctuations are large. There seems to be no systematic shift between the distribution 
of differences of red (metal-rich) and blue (metal-poor) peaks for GCS with 
bimodal metallicity distribution. 

\begin{figure}[tbp]
\resizebox{\hsize}{!}{\includegraphics*{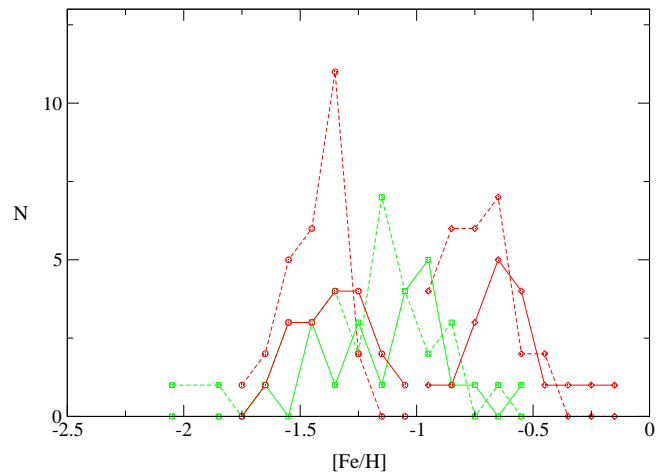}}
\caption{Comparison of metallicities of GCS, determined by the present study (solid lines) 
and by previous studies (dashed lines). Squares -- unimodal systems, 
circles -- the metal-poor component of bimodal systems, diamonds -- the metal-rich 
component of bimodal systems.}
\label{f10label}
\end{figure}

\begin{figure}[tbp]
\resizebox{\hsize}{!}{\includegraphics*{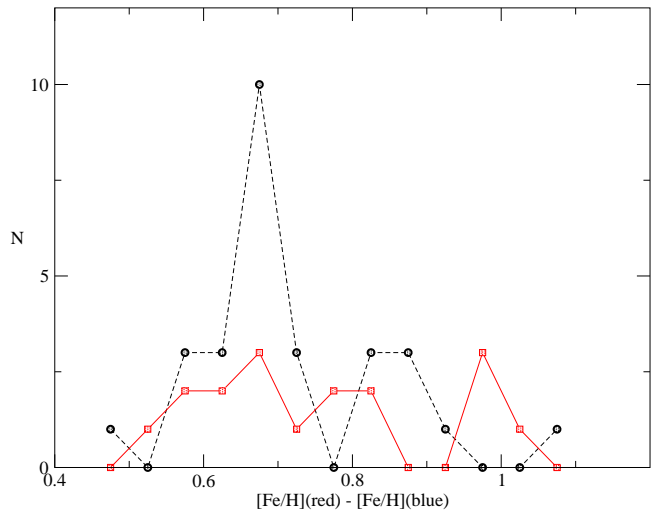}}
\caption{Comparison of the distribution of differences between red (metal-rich) 
and blue (metal-poor) peaks for GCS with bimodal metallicity distribution, 
determined by the present study (solid lines) and by previous studies (dashed lines).}
\label{f11label}
\end{figure}

\begin{figure}[tbp]
\resizebox{\hsize}{!}{\includegraphics*{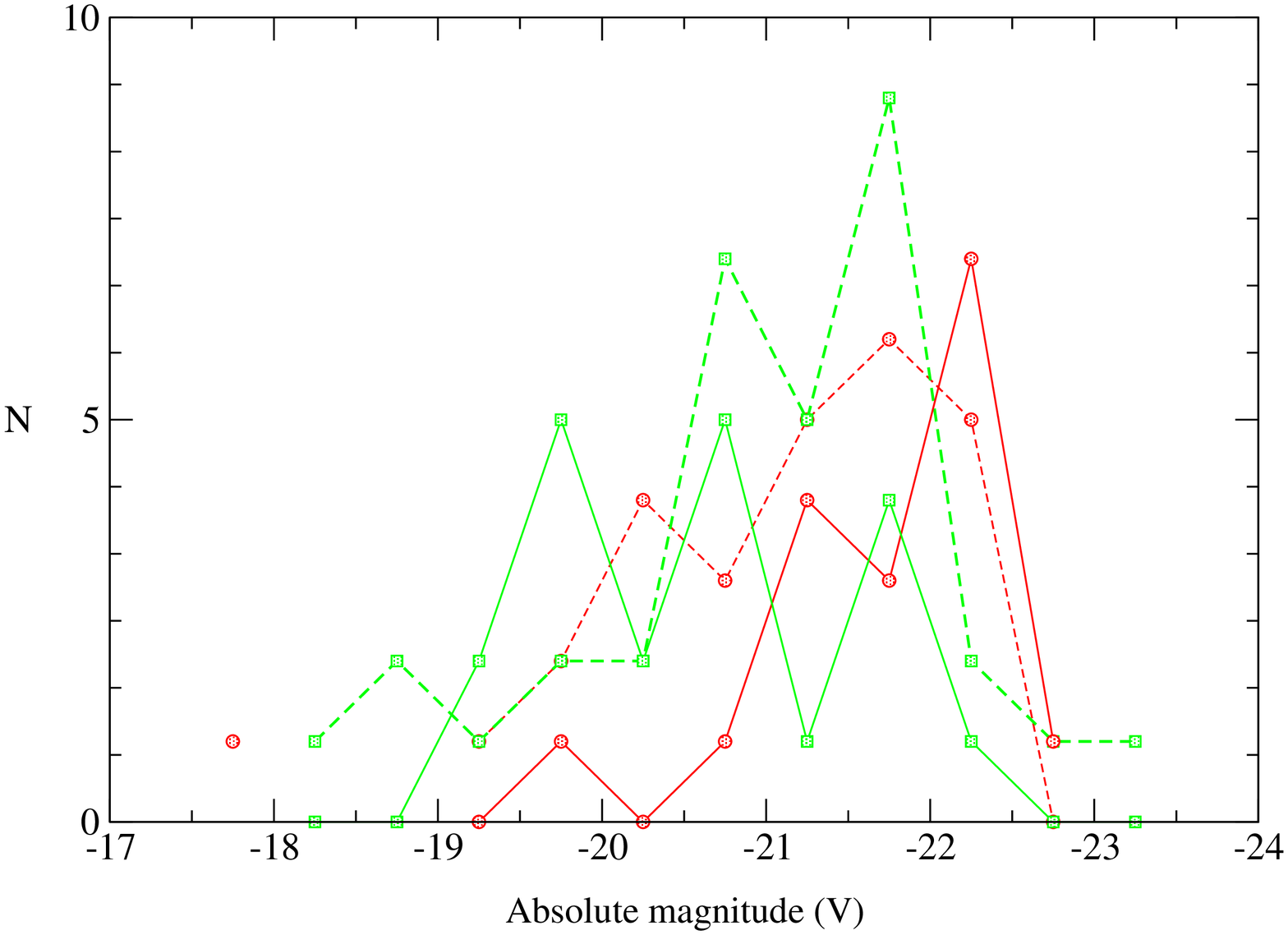}}
\caption{Comparison of the distribution of absolute magnitudes of galaxies, 
determined by the present study (solid lines) and by previous studies (dashed lines). 
Circles -- galaxies with bimodal metallicity distribution of GCs, squares -- galaxies 
with unimodal metallicity distribution of GCs.}
\label{f12label}
\end{figure}

Next, we compare the metallicities of GCSs derived in the present study to those 
by other authors. We would like to stress that the metallicities are in both cases 
determined with help of the KMM test. In Figs.~\ref{f10label}-\ref{f12label} we repeate 
Figs.~\ref{f5label}-\ref{f7label} but separately for two samples. Solid lines correspond 
to the sample of GCSs, metallicities of which are determined in the present study,
dashed lines correspond to the sample of other papers. We think that there are no 
systematic differences between the two samples.

\begin{table}[ht]
\caption{Metallicity peaks for bimodal distributions}
\label{bimodal}
\begin{tabular}{lllll}
\hline
Galaxy& Refer. &$\rm [Fe/H]_1$ & $\rm [Fe/H]_2$ & Source\\ 
\hline
224   & 22 & -1.43 & -0.60 &  22 \\ 
524   & 1  & -1.30 & -0.61 &  1  \\  
584   & 2  & -1.30 & -0.70 &  2  \\ 
1023  & 3  & -1.46 & -0.61 &  3  \\ 
1052  & 4  & -1.56 & -0.97 &  4  \\ 
1199  & 8  & -1.49 & -0.84 &  8  \\ 
1380  & 5  & -1.66 & -0.71 &  *  \\ 
1399  & 23 & -1.22 & -0.30 &  *  \\ 
1404  & 6  & -1.45 & -0.67 &  *  \\ 
1427  & 7  & -1.33 & -0.67 &  7  \\ 
1439  & 8  & -1.33 & -0.71 &  8  \\ 
1700  & 9  & -1.56 & -0.84 &  *  \\ 
2768  & 8  & -1.52 & -0.87 &  8  \\ 
3115  & 1  & -1.36 & -0.67 &  *  \\ 
3311  & 10 & -1.52 & -0.94 &  *  \\ 
3377  & 8  & -1.36 & -0.80 &  8  \\ 
3379  & 1  & -1.36 & -0.67 &  1  \\ 
3384  & 1  & -1.43 & -0.54 &  1  \\ 
3923  & 11 & -1.23 & -0.58 &  *  \\ 
4278  & 8  & -1.46 & -0.80 &  8  \\ 
4365  & 1  & -1.29 & -0.63 &  1  \\ 
4406  & 1  & -1.28 & -0.76 &  *  \\ 
4472  & 12 & -1.10 & -0.50 &  *  \\ 
4473  & 1  & -1.44 & -0.72 &  1  \\ 
4478  & 13 & -1.26 & -0.28 &  *  \\ 
4486B & 2  & -1.52 & -0.90 &  2  \\ 
4486  & 14 & -1.39 & -0.59 &  *  \\ 
4494  & 1  & -1.55 & -0.89 &  1  \\ 
4526  & 2  & -1.62 & -0.74 &  2  \\ 
4552  & 1  & -1.39 & -0.67 &  1  \\ 
4565  & 15 & -1.05 & -0.40 &  *  \\ 
4594  & 16 & -1.42 & -0.63 &  *  \\ 
4621  & 13 & -1.30 & -0.71 &  *  \\ 
4649  & 1  & -1.38 & -0.56 &  1  \\ 
4660  & 13 & -1.46 & -0.97 &  13 \\ 
5128  & 17 & -1.11 & -0.16 &  *  \\ 
5846  & 18 & -1.36 & -0.67 &  *  \\ 
5982  & 19 & -1.30 & -0.65 &  19 \\ 
6702  & 20 & -1.79 & -0.71 &  20 \\ 
6868  & 8  & -1.52 & -0.84 &  8  \\ 
7332  & 8  & -1.62 & -0.97 &  8  \\ 
7619  & 8  & -1.29 & -0.45 &  8  \\ 
IC1459& 19 & -1.30 & -0.48 &  19 \\ 
IC4051& 21 & -1.30 & -0.71 &  21 \\ 
MW    & 22 & -1.59 & -0.55 &  *  \\ 
\hline
\end{tabular}
\\
References for GC colour determinations: 1 -- Larsen et al. (2001a); 2 -- 
Gebhardt \& Kissler-Patig (1999); 3 -- Larsen \& Brodie (2000); 4 -- Forbes
et al. (2001); 5 -- Kissler-Patig et al. (1997b); 6 -- Forbes et al. (1998);
7 -- Forte et al. (2001); 8 -- Forbes (2001); 9 -- Brown et al. (2000); 10 -- 
Brodie et al. (2000); 11 -- Zepf et al. (1995); 12 -- Geisler et al. (1996);
13 -- Neilsen \& Tsvetanov (1999); 14 -- Cohen et al. (1998); 15 -- 
Kissler-Patig et al. (1999); 16 -- Larsen et al. (2001b); 17 -- Geisler \& Forte
(1990); 18 -- Forbes et al. (1997); 19 -- Forbes et al. (1996); 20 -- 
Georgakakis et al. (2001); 21 -- Woodworth \& Harris (2000); 22 -- Barmby et al. 
(2000); 23 -- Ostrov et al. (1998). Sources for bimodality determinations are 
the same with the exception: * -- present study.
For galaxies 1380, 1399, 1404, 1427, 3923, 6702 reddening correction is taken from 
the model by Burstein \& Heiles (1984).
\end{table}

In a subsequent paper we shall study correlations of several GCS properties with 
the parameters of galaxies and galactic environment in more detail.

\begin{table}[ht]
\caption{Metallicity peaks for unimodal distributions}
\label{unimodal}
\begin{tabular}{llll}
\hline
Galaxy& Ref. &$\rm [Fe/H]$ & Source\\ 
\hline
474  & 1  & -1.33 & 1 \\ 
596  & 2  & -1.07?& * \\ 
598  & 10 & -1.60 & * \\ 
821  & 2  & -0.97?& * \\ 
1201 & 1  & -1.10?& 1 \\ 
1332 & 1  & -1.00?& 1 \\ 
1374 & 3  & -0.98 & * \\ 
1379 & 3  & -0.70 & * \\ 
1387 & 3  & -0.50 & * \\ 
1389 & 1  & -1.30 & 1 \\ 
1400 & 1,4& -1.03 & 1 \\ 
1426 & 2  & -0.96?& * \\ 
1553 & 1  & -1.13 & 1 \\ 
2300 & 2  & -1.00 & * \\ 
2434 & 2  & -1.41?& * \\ 
2778 & 2  & -0.99 & * \\ 
2902 & 1  & -0.67 & 1 \\ 
3031 & 11 & -1.48 & * \\ 
3056 & 1  & -1.00 & 1 \\ 
3414 & 1  & -0.94 & 1 \\ 
3489 & 1  & -1.43?& 1 \\ 
3585 & 2  & -0.87 & * \\ 
3599 & 1  & -1.52 & 1 \\ 
3607 & 1  & -1.16 & 1 \\ 
3608 & 5  & -1.25 & * \\ 
3610 & 2  & -1.10 & 2 \\ 
3640 & 2  & -0.97?& * \\ 
4125 & 2  & -1.23 & * \\ 
4192 & 2  & -1.61 & 2 \\ 
4203 & 1  & -1.20 & 1 \\ 
4291 & 2  & -1.30 & 2 \\ 
4343 & 2  & -1.14 & 2 \\ 
4374 & 2  & -1.43 & * \\ 
4379 & 1  & -1.39 & * \\ 
4450 & 2  & -1.42 & 2 \\ 
4458 & 6  & -1.13?& * \\ 
4459 & 1  & -1.16 & 1 \\ 
4536 & 2  & -2.00 & 2 \\ 
4550 & 2  & -1.30?& 2 \\ 
4569 & 2  & -1.83 & 2 \\ 
4589 & 5  & -1.10 & 5 \\ 
4874 & 7  & -1.53 & 7 \\ 
4881 & 2  & -1.51 & 2 \\ 
5018 & 2  & -1.46 & 2 \\ 
5061 & 2  & -0.91 & 2 \\ 
5322 & 5  & -1.07 & 5 \\ 
5813 & 5  & -1.20 & 5 \\
5845 & 2  & -1.10 & 2 \\
5907 & 8  & -1.24 & * \\ 
6703 & 1  & -0.80 & 1 \\ 
6861 & 1  & -0.87?& 1 \\ 
7192 & 2  & -1.01 & * \\ 
7457 & 9  & -1.09 & * \\ 
7626 & 5  & -0.84 & 5 \\ 
\hline
\end{tabular}
\\
References for GC colour determinations: 1 -- Kundu \& Whitmore (2001a);
2 -- Gebhart \& Kissler-Patig (1999); 3 -- Kissler-Patig et al. (1997a); 4 --
Perrett et al. (1997); 5 -- Forbes et al. (1996); 6 -- Neilsen \& Tsvetanov
(1999); 7 -- Harris et al. (2000); 8 -- Kissler-Patig et al. (1999);
9 -- Chapelon et al. (1999); 10 -- Sarajedini et al. (1998); 11 -- Perelmuter
(1995). Sources for unimodality determinations are the same with the exception:
* -- present study.
Metallicities with ``?'' indicate rather wide distribution.
Reddening in a study by Kundu \& Whitmore (2001a) is taken according to
Burstein \& Heiles (1984).
\end{table}

\acknowledgements

We would like to thank the anonymous referee for useful comments and suggestions.
We acknowledge the financial support from the Estonian Science Foundation 
(grant 4702) and DAAD (grant A0209036). Part of the paper was written at 
Goettingen University Observatory and we thank the staff for hospitality.
We also thank P. Barmby, D. Forbes, K. Gebhardt, J. Kavelaars, S. Larsen, 
Z. Tsvetanov and S. Zepf for communicating their data in numerical form as 
electronic files.

\end{document}